\documentclass[aip,reprint,numerical,a4paper,superscriptaddress]{revtex4-1}

\usepackage{graphicx,natbib}
\usepackage{amsmath,amssymb,amsfonts,bm}
\usepackage[T1]{fontenc}
\usepackage[usenames,dvipsnames]{xcolor}
\usepackage{soul}
\usepackage{xfrac}
\usepackage{natbib}
\usepackage[maxfloats=25]{morefloats}

\newcommand{\ppcode}{\texttt{Photon-Plasma }code}

\newcommand{\owpe}{\omega_{pe}^{-1}}

\newcommand{\dx}{\Delta x}

\newcommand{\pp}{\bm{p}}
\newcommand{\JJ}{\bm{J}}
\newcommand{\BB}{\bm{B}}

\newcommand{\rr}{\bm{r}}
\newcommand{\vv}{\bm{v}}

\newcommand*\slfrac[2]{{}^{#1}\!/_{#2}}

\begin{document}

\title{Particle Control in Phase Space by Global K-Means Clustering}

%--------------------------------------------------------
\author{J.~Trier Frederiksen} 
\affiliation{Niels Bohr International Academy, Niels Bohr Institute, Blegdamsvej 17, 2100 K\o benhavn \O, Denmark}
\email{trier@nbi.dk}
\author{G.~Lapenta}
\affiliation{Centre for mathematical Plasma Astrophysics, Department of Mathematics,
KU Leuven, Celestijnenlaan 200B, 3001 Heverlee, Belgium}
\author{M.~Pessah}
\affiliation{Niels Bohr International Academy, Niels Bohr Institute, Blegdamsvej 17, 2100 K\o benhavn \O, Denmark}

%-----------------------------------------
\begin{abstract}
\noindent We devise and explore an iterative optimization procedure for controlling particle populations in particle-in-cell (PIC) codes via merging and splitting of computational macro-particles. 
Our approach, is to compute an optimal representation of the global particle phase space structure while decreasing or increasing the entire particle population, based on k-means clustering of the data. In essence the procedure amounts to merging or splitting particles by statistical means, throughout the entire simulation volume in question, while minimizing a 6-dimensional total distance measure to preserve the physics. 
Particle merging is by far the most demanding procedure when considering conservation laws of physics; it amounts to lossy compression of particle phase space data. We demonstrate that our k-means approach conserves energy and momentum to high accuracy, even for high compression ratios, $\mathcal{R} \approx 3$ --- \emph{i.e.}, $N_{f} \lesssim 0.33N_{i}$. 
Interestingly, we find that an accurate particle splitting step can be performed using k-means as well; this from an argument of symmetry. The split solution, using k-means, places splitted particles optimally, to obtain maximal spanning on the phase space manifold.
Implementation and testing is done using an electromagnetic PIC code, the \ppcode. Nonetheless, the k-means framework is general; it is not limited to Vlasov-Maxwell type PIC codes. We discuss advantages and drawbacks of this optimal phase space reconstruction.  \\
\end{abstract}

%-----------------------------------------%
                                   % howto put PACS?

\maketitle

%========================================================================
\section{Introduction} \label{sec:introduction}

Control of computational macro-particle (CMP) populations in Particle-In-Cell (PIC) codes is particularly desirable in at least two situations:
\begin{description}
 \item[Population Runaway] Monte Carlo realizations of collisional processes in PIC codes, for example, often involves fractionation of CMPs into ''parents'' and ''children'' for enhanced statistical resolution of the collision processes. This results in explosion of CMP populations, and a memory bounded simulation longevity. 
 \item [Load balancing] in PIC codes relies on the ability to redistribute CMPs among computational processes (e.g. in MPI domain decomposed models) at runtime to maintain similar execution times of the computational processes, and preserve statistical resolution of continuous phase space.
\end{description}

CMP de-population (re-population) of domains that are progressively filled (depleted) can be achieved through deletion (addition) of CMPs -- for those domains which are oversampled (undersampled), while attempting to maintain physical quantities locally conserved. Single particle deletion (addition) CMPs is detrimental with respect to the conservation of the physical properties of the system being modeled~\cite{lapenta1,lapenta2,lapenta3,scottmartin1,grasso1}. It is necessary to merge (split) several CMPs to conserve both momentum and energy from the phase space information available. \\

\noindent An algorithm that can achieve this goal in a robust and efficient manner will benefit a wide range of problems in laboratory and astrophysical settings. Many physical processes naturally lead to runaway CMP populations (time domain), and extreme CMP concentrations (spatial domain), e.g.

\begin{figure}[!h]
\begin{center}
 \includegraphics[width=\columnwidth]{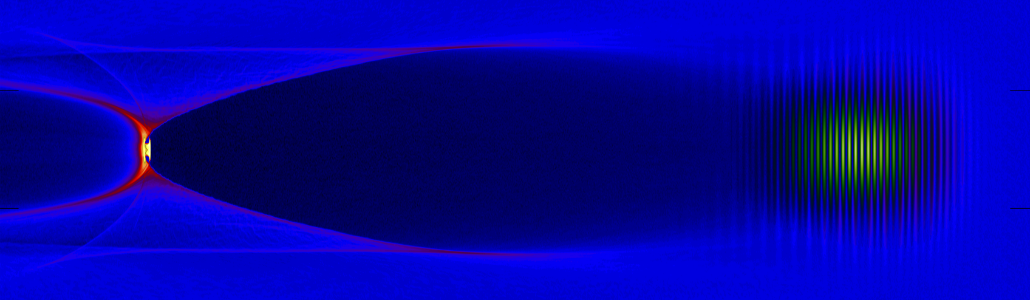}
\caption{Early stage in ultra-high intensity laser pulse interacting with a quiescent homogenous plasma plume, showing electron CMP number density (colors not to scale). A bubble (dark central region), evacuated of electrons, is created by the highly non-linear disturbance from the laser field pulse (green, right), which is propagating to right. A hot-spot (yellow, left) in the wake of the laser pulls electrons along, at close to the speed of light. The highly inhomogenous density, ranging from $N_{\mathrm{e,min}}\approx1$ to $N_{\mathrm{e,max}}\approx300$ severely affects load balancing. Our method can alleviate this problem to speed up the simulation by a significant factor.}
\label{fig:laserplasma_interaction}
\end{center}
\end{figure}

\begin{description}
\item [Load] High-intensity laser-plasma wakefield acceleration of electrons, \citet{beck1} (also Figure~\ref{fig:laserplasma_interaction}).
\item [Runaway] Gamma-Ray Burst wakefield plasma acceleration, under the influence of detailed Compton scattering, \citet{frederiksen1}.
\item [Load] Streaming instabilities and agglomeration of planetesimals leading to planet formation, \citet{johansen1}.
\item [Runaway \& load] High-energy radiative processes and pair cascades in pulsar magnetospheres, \citet{timokhin1}.
\item [Load] Streams and caustics in the evolution of dark matter structures in cosmological simulations, \citet{vogelsberger1}.
\end{description}

All these cases (and many others) demand an efficient CMP population control and/or redistribution in large-scale numerical simulations. \\

\noindent Several strategies for CMP merging have been visited in the literature over the last few decades, changing in order of complexity, cost and accuracy. ~\citet{lapenta1}, also later~\citet{lapenta3} and \citet{teunissen1}, considered the problem of merging/splitting on a single particle basis, e.g., $2\leftrightarrow1$, $3\leftrightarrow2$, and cell-based $N_{cell}\leftrightarrow M_{cell}$ approaches, (\citet{lapenta2}), with $N$ and $M$ small. More recently, more complex algorithms have emerged such as agglomerate clustering~\cite{grasso1} and resampling, and also oct-tree reconstruction~\cite{scottmartin1} in momentum space.

Commonly, those previous strategies used means of algebraic reconstruction to ensure that physical field quantities, represented on the PIC discrete mesh (in $\rr$-space) would be conserved exactly. Some were investigated in reduced-dimensional systems, e.g. 1D3V (\citet{scottmartin1}), although their method was not strictly constrained to 1D. Others further applied a reconstruction procedure which decomposed 6D phase space, $f(\rr,\pp,t)$, in to 3D subspaces, $f_r(\rr,t)$ and $f_p(\pp,t)$, employing strict algebraic reconstruction on $\rr$-space, while retaining the solution found by agglomerate clustering in $\pp$-space (\citet{grasso1}). Any decomposition of phase space, $\mathbb{R}^\textrm{D}$, into phase subspaces $\mathbb{R}^\textrm{B}$ and $\mathbb{R}^\textrm{C}$, with B+C=D (for our case D=6), removes information contained in possible cross-correlation between the subspaces. It is conceivable that such correlations should be preserved.

In a view alternative to previous strategies, we consider the problem of reducing (increasing) particle phase space resolution by merging (splitting) CMPs, as an optimization problem in 6 dimensions. Our approach randomly selects existing particles as a global best guess at a solution for the clustering, with the objective to either merge or split them into a new imitative set of particles. Subsequently, a K-means iterative minimization of a global intra-cluster distance measure successively drives the merged (split) solution towards a reduced (increased) CMP population, with the same physical properties. \\

\noindent In Section~\ref{sec:kmeans}, we describe the natural relationship between k-means clustering and the PIC code phase space representation. We then describe the details of our global k-means procedure; initialization, distance measure, particle merging and splitting, as well a crude, yet important, edge preserving measure to circumvent k-means artifacts on bounded domain decompositions. Section~\ref{sec:barbara} outlines our test simulation setup and presents a few crucial tests of our k-means clustering procedure. Discussion and conclusions are given in Section~\ref{sec:discussion}.

%===================================================================
\section{K-Means Clustering in the PIC Codes} \label{sec:kmeans}
Generally, in electromagnetic PIC codes, the source terms in Maxwell's equations, $\rho_c(\rr,t)$ and $\JJ(\rr,\pp,t)$, are constructed from interpolated accumulation of a large number of computational macro-particles (CMPs) onto a computational mesh. These CMPs are distributed in \emph{continuous} real space and momentum space, and given a continuous weight to signify the particle statistical influence. For very large numbers of CMPs, we can approximately describe the computational plasma everywhere by a distribution function, $f(\rr,\pp,t)\equiv\sum_sf_s(\rr,\pp,t)$, hereafter phase space density, where the subscript '$s$' denotes particle species.

In the \ppcode\cite{ppcode_paper}, for the most complete case of 3D3V simulations, the CMP is represented by a six-tuplet of real numbers, ${\widetilde{\rr}}\equiv\{r_x,r_y,r_z,p_x,p_y,p_z\}$, which positions the particle in 6-dimensional phase space (the tilde signifies a 6D vector). Further each CMP is given a statistical weight, $w_i$, which dictates a relative strength of the particle with respect to either the number of physical particles, or a scaled amount of physical particles. Relativistic momentum is $\pp \equiv m_0\gamma(v)\vv$, with $\gamma \equiv \sqrt{(1 - \beta^2)}$, $\beta \equiv v/c$, and in the \ppcode~ we most naturally keep the CMPs' relativistic 3-velocity, $\pp/m_0$. For example $p_{z}=v_{z}(1-\beta^{2}_{z})^{-1/2}$ with $\beta_{z} \equiv v_{z}^2/c^2$. This renders direct addition and subtraction of particle momenta, vectorially, physically meaningful. Consequently, we may view the particle ensemble phase space as a collection of points in 6-dimensional Euclidian affine space, with a well defined algebra consisting of addition (e.g. $p_{z,i} + p_{z,j} = p_{z,k}$), subtraction (e.g. $x_i-x_j=x_k$) and a distance measure, 
\begin{equation}
d^2(\widetilde{\rr}_i,\widetilde{\rr}_j)=(x_i-x_j)^2 + \ldots + (p_{z,i}-p_{z,j})^2.
\end{equation}
Particles can then be vectorially added or subtracted, and we can find a distance between them in this affine space. We can also construct an arithmetic mean, or for weighted particles, a weighted arithmetic mean of any ensemble, or cluster center point, of particles 
\begin{equation}
\label{eq:weighted_arithmetic_mean}
\widetilde{\overline{\rr}} = \frac{\sum_i w_i\widetilde{\rr}_i}{\sum_i w_i} ~~ \Leftrightarrow ~~\widetilde{\overline{\rr}} w_{cl} = \sum_i w_i\widetilde{\rr}_i~~,~ w_{cl} \equiv \sum_i w_i,
\end{equation}
where now barred vectors, i.e. $\widetilde{\overline{\rr}}$, denotes cluster points. \\

\noindent These simple facts form the basis of this paper and the justification of global k-means clustering as a way of optimal phase space reconstruction in, for example, PIC codes.

%----------------------------------------------
\subsection{Weighted k-means clustering}\label{sec:weighted_kmeans}
Multivariate, multidimensional, data can be analyzed and manipulated using vector compression. K-Means belongs to this general class of vector compression algorithms\cite{jmacqueen1}, and can be used to either refine or coarsen multivariate data manifolds. In this article, the \underline{weighted} k-means\cite{chan2004} objective is: from a set of M data points, $\{\widetilde{\rr}_1,...,\widetilde{\rr}_M\}$, with weights $\{w_1,...,w_M\}$, in D-dimensional space, $\mathbb{R}^D$, find K cluster centers, $\{\widetilde{\overline{\rr}}_1,...,\widetilde{\overline{\rr}}_K\}$, with weights $w_1,...,w_K$, also in $\mathbb{R}^D$, which partition the original data in the optimal way. This is defined as that partitioning which minimizes the total global intra-cluster distance, 
\begin{equation}
\label{eq:kmeansobjective}
\texttt{min}(\mathcal{\widetilde{D}}_{tot}) \equiv \texttt{min}\left(\sum_{j=1}^K \sum_{\widetilde{\rr}_i \in \widetilde{\overline{\rr}}_j} w_i \| \widetilde{\rr}_i - \widetilde{\overline{\rr}}_j\|^2\right)~,
\end{equation}
with $\widetilde{\overline{\rr}}_j$ defined as the j'th cluster center by equation~\ref{eq:weighted_arithmetic_mean} (left), respectively (right).  \\

\noindent We choose to work in this paper in normalized data space, such that 
$\{r_x, ..., p_z\} \rightarrow \{r_x/L_{x}, ..., p_z/L_{pz}\}$, where 
$\{L_{x}, ..., L_{pz}\} \equiv \{ \mathtt{max}(r_x) - \mathtt{min}(r_x), ..., \mathtt{max}(p_z)-\mathtt{min}(p_z)\}$. We cannot \emph{a priori} assume that certain directions in phase space are more important than others with respect to the physics, if we want the procedure to be generally applicable. 

We did test the k-means procedure also, using non-normalized data space, i.e. $\{r_x, ..., p_z\}$, to see how this would affect the merged solution, for the case of a thermal plasma. Significant differences were found between the solutions in the two very different representations of the phase space data. We give a few results, superficially, in the section on tests (below). 

The choice of a distance norm, and the choice of normalization, of the data space severely impacts the quality of the k-means solution. Our choice of normalized data spaces should be the general one, not to favor certain directions in phase space. However, it is beyond the scope of this paper to investigate the details of such choices, as concerning distance measures and normalization.

~ \\ ~ \noindent In signal compression theory, the original data set to be compressed or inflated in k-means is often denoted 'training vectors' while the solution (the clustered data set) is called the 'codebook vectors'. We adopt this terminology henceforth.

\begin{figure}[!h]
\begin{center}
 \includegraphics[width=\columnwidth]{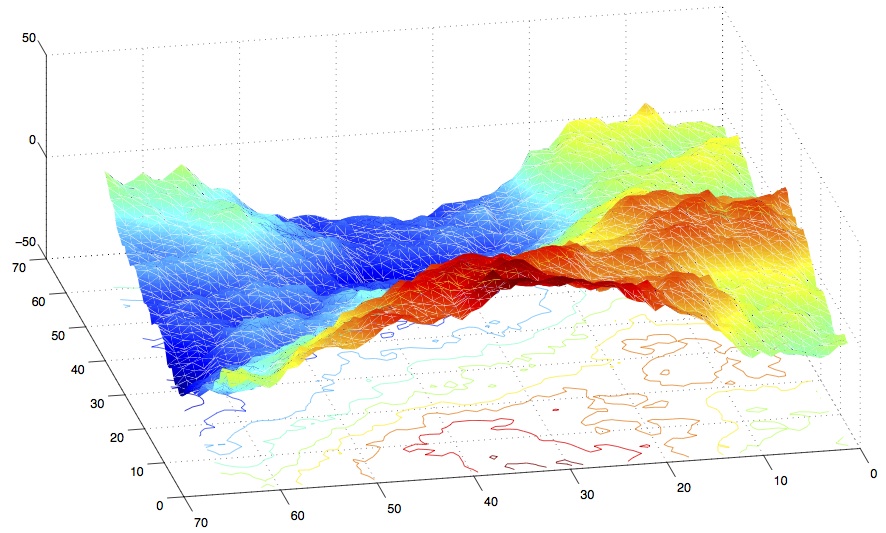}
\caption{Illustration: finding the global minimum in k-means is NP-hard for $D\geq2$ dimensions and $K\geq2$. However, for approximate solutions, i.e. when finding only a \emph{sufficiently} global minimum, heuristic algorithms converge quickly. Arbitrary data set generated in MATLAB\cite{lakaemper1}.}
\label{fig:landscape}
\end{center}
\end{figure}

Finding the \emph{global} minimum for any data set in higher dimensions in k-means is an NP-hard task. For given values of $M$, $K$ and $D$, the computational effort is approximately $\mathcal{O}(M^{KD+1}logM)$ which is intractable for almost any PIC code problem we want to consider. If we --- on the other hand --- accept the solution to be only approximative we can find acceptable alternatives in finite time, and even quite fast. Equivalently, an approximate solution amounts to a \emph{local} minimum rather than the global minimum described by Equation~\ref{eq:kmeansobjective}).

%----------------------------------------------
\subsubsection{K-Means Clustering, Lloyd-Forgy algorithm}
\begin{figure}[!h]
\begin{center}
 \includegraphics[width=.75\columnwidth]{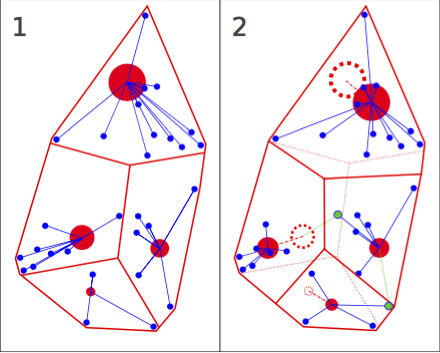}
\caption{Weighted Lloyd-Forgy iterative clustering (''k-means'') in 2D. \emph{Panel 1}: initialize codebook (red disks) at randomly chosen training vectors (blue dots), and perform Voronoi tesselation. \emph{Panel 2}: compute new weighted arithmetic means of training vectors within each Voronoi cell and re-define these means as the codebook vectors. Adjust weights. NB: some training vectors will migrate to other cells (green dots). Dashed lines represent previous iteration; Voronoi cell boundaries (red lines), cluster positions (red open circles) and migrated training vectors (green lines). Reiteration is performed until a convergence criterion is met.}
\label{fig:kmeans_sketch}
\end{center}
\end{figure}

A variety of heuristic algorithms exist; commonly they use iterative processes to find a local minimum solution to Equation~\ref{eq:kmeansobjective}. The simplest brute force heuristic algorithm, which is also the most expensive, is Lloyd's algorithm\footnote{Formally, the original formulation considered only evenly spaced points in one dimension, but the algorithm is not limited to specific point data densities nor one dimension.} \citet{lloyd1} with Forgy initial conditions \citet{forgy1}. We will use ''Lloyd's'' algorithm and ''k-means'' interchangably, even though the ''k-means'' term and a more general treatment of vector quantization originates from~\citet{jmacqueen1}.  Lloyd-Forgy, or k-means clustering optimization goes through three basic steps:

\begin{description}
\item[1. Initial condition] a first guess as to a solution is made by placing the initial codebook vector set. Forgy's method at random selects K training vectors as the initial codebook. This often (but not always) is better than for example choosing random points within the data space.
\item[2. Cluster assignment] training vectors are assigned each to their nearest codebook vector (cluster center). This is effectively a Voronoi tesselation step.
\item[3. codebook replacement] by calculating the weighted arithmetic mean (Equation~\ref{eq:weighted_arithmetic_mean}), based on within-cluster associated training vectors, new codebook centers are found to replace those codebook vectors found in 2) during the previous iteration.
\end{description}
Steps 2 and 3 are repeated until some defined convergence threshold is met; for example, as in this paper, when the ratio in total error (eqn.~\ref{eq:kmeansobjective}) between successive iterations changes by less than 1.0\% is a common criterion. Figure~\ref{fig:kmeans_sketch} illustrates the algorithm for a two-dimensional case.

The effect of successively tessellating and cluster re-centering, respectively, reduces the computational effort to $\mathcal{O}(M \times K \times D \times i)$, with $i$ the number of iterations to convergence. Nonetheless, even when employing Lloyd-Forgy, the computational expense becomes increasingly prohibitive for large values of $M$, $K$ and $D$. Hence, we might expect to discard k-means as feasible for CMP merging/splitting in PIC codes, especially for global or semi-global simulation volumes. In this paper we demonstrate its feasibility in terms of \emph{physics}, rather than consider computational feasibility. Elsewhere (\citet{their_paper}) it is reported that by employing various accelerated partitioning and distance calculations, or employing brute force GPGPU kernels, the running time is reduced to acceptable levels, thus demonstrating its feasibility in terms of \emph{computation} as well.

%----------------------------------------------
\subsubsection{Particle merging --- employing k-means}
From the previous section, merging particles many-to-many, globally (or semi-globally) in the volume now becomes obvious; after the k-means operation, the codebook will contain all the necessary phase space information needed to preserve the physics in the continued simulation. 

We only need to delete the original particle data (the training vectors) and replace them with the new reduced particle data set (the codebook) 
\begin{equation}
\{\widetilde{\rr}_{1},...,\widetilde{\rr}_{M}\}_{\{s,tr\}} \rightarrow \{\widetilde{\rr}_{1},...,\widetilde{\rr}_{K}\}_{\{s,cb\}}~,
\end{equation}
while conserving total charge, globally, by preserving the total weight of the CMPs, pre- and post-compression:
\begin{equation} \label{eq:weight_conserve}
\sum_j^Kw_{cb} = \sum_i^Mw_{tr}~,
\end{equation}
One further constraint is
\begin{equation} \label{eq:weight_conserve}
w_{cb} = \sum_l^{N_{cl}}w_{tr,l}~,
\end{equation}
for all $N_{cl}$ intra-cluster particles. Here 'cb' ('tr') denoting codebook (training) vectors, respectively, and 's' denoting species.

Several schemes exploit the additive properties of phase space, and they can be classified according to the approaches mentioned in the Introduction. The advantage of a many-to-many ($M \rightarrow K$, $M > K >> 1$) iterative optimizing approach, like ours, is that we do not have to consider specifically, nor analytically, conservation of energy, momentum, space charge density, current density or any higher order moments of the distribution. Many degrees of freedom make it possible to satisfy conservation laws of physics to high precision\footnote{We do not need an exact solution, only one good enough to subside the Poisson noise in the original CMP ensemble.}. The quality of the iterated solution will however be practically constrained by computational expense, and by demands on the number of particles in the simulation.

%----------------------------------------------
\subsubsection*{Energy and momentum conservation}\label{sec:e_p_cons}
In $D$ dimensions, a particle has $D$ degrees of freedom. Momentum conservation demands, then, $D$ constraints and energy conservation an additional constraint, for a total of $D+1$ constraints. Consequently, in any dimensionality $D$, when merging $M$ particles into $K$ particles, the resulting particle number ($K$) must be strictly larger than one, to supply the needed degrees of freedom for simultanous momentum and energy conservation.

\begin{figure}[!h]
\begin{center}
\includegraphics[width=0.8\columnwidth]{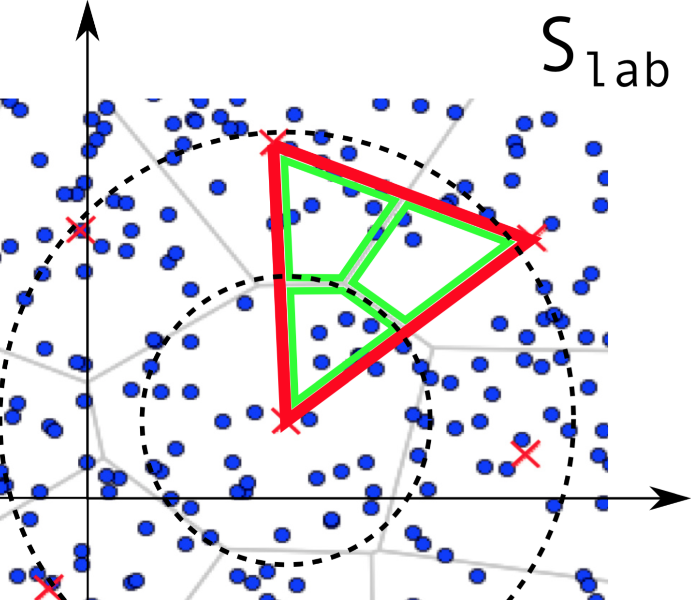}
\caption{The Voronoi-/Delaunay-cell dual, and their interpretation in terms of training vector set, and code book vector solution. \emph{Blue dots:} original training vectors, \emph{red crosses:} solution (merged) code book vectors. \emph{Grey lines:} Voronoi-cell boundaries, \emph{thick red lines:} Delaunay-cell boundaries, \emph{green lines:} individual Delaunay-cells' share of training vector set enclosed in Delaunay-cell volume. \emph{Black dashed circles:} integration over shells around a Voronoi-cell center (cluster center) for energy conservation test.}
\label{fig:VoronoiDelaunay}
\end{center}
\end{figure}

The fundamental cluster (unit cell) in a k-means solution is a Voronoi-cell, by tesselation. A dual set exists, which is the Delaunay-triangulation --- the Delaunay-cell. This dual relation is sketched\footnote{We restrict the discussion to a 2-dimensional phase space for illustration; the argument is general to $D$ (6) dimensions.} in Figure~\ref{fig:VoronoiDelaunay}.

While it is clear that a single Voronoi-cell cannot conserve both momentum and energy, simply because the single cluster particle does not have sufficient degrees of freedom\footnote{This is true for \emph{all} merging schemes.}, it is equally clear from Figure~\ref{fig:VoronoiDelaunay} that the Delaunay-cell has sufficient degrees of freedom to provide simultaneous conservation of momentum and energy, to the highest possible accuracy. The Delaunay-cell, which is constituted by three (in 2D) cluster center particles (codebook vectors) carries the share of phase space information from their respective partial neighboring Voronoi-cells. 

Further, any Voronoi-cell center (cluster center) is the rest frame of any cell calculation, by k-means construction, namely it is the weigthed arithmetic average of cluster members. Again --- by construction --- therefore its momentum vanishes in that (local) frame, 
\begin{equation}
\mathbf{P}^2_{cl} = \frac{1}{\bar{w}^2_{cl}} \sum_l w_l\mathbf{p}^2_l \equiv 0~,
\end{equation}
to convergence criterion accuracy. The Voronoi-cell cluster member particles (training vectors) all have non-zero momentum (therefore energy) in this restframe, i.e. 
\begin{equation}
E_{cl} = \frac{1}{\bar{w}^2_{cl}} \sum_l w_l E_l = \sum_l w_l \sqrt{\mathbf{p}^2_l + m^2_0} ~~ \rm{(c \equiv 1)} ~.
\end{equation}
Merging the cluster members onto cluster centers will delete local Voronoi-cell information about the differential energy contributions, namely the terms $\mathbf{p}_l$ in the expression above. The result is a loss of energy from the cluster particle --- not a loss of rest mass (text below).

The cluster rest mass, 
\begin{align}\label{eq:clusterrestmass}
 E^2_{cl} - \mathbf{P}^2_{cl} &\equiv M^2_{0,l} \neq \sum_l m^2_{0,l} \neq  \nonumber \\
 & \frac{1}{\bar{w}^2_{cl}} \left[ \left( \sum_l w_l \sqrt{\mathbf{p}^2_l + m^2_0} \right)^2 - \left( \sum_l w_l \mathbf{p}_l \right)^2 \right] ~,
\end{align}
therefore generally contains the error introduced by loosing the relativistic energy contribution --- due to the merge --- from the '$l$' cluster members, $\mathbf{p}^2_l$), under the square root in Equation~\ref{eq:clusterrestmass}\footnote{We sincerely thank one of our referees for discussion on this aspect of energy and momentum conservation. One of our referees' suggestions have lead to significant improvement of this section.}. 

Relativistic energy is conserved but not invariant, whereas rest mass is invariant but not conserved. We therefore interpret the error in relativistic energy, not as relativistic mass, and the total rest mass should not change.

The contribution will be small since it is a local rest frame contribution, but it cannot be avoided when conducting lossy data compression (a merge); there will always be an error term when merging particles. This error in rest frame energy is what the k-means algorithm minimizes (in conjunction with spatial position). 

We can now concretely define what the convergence objective, Equation~\ref{eq:kmeansobjective}, means. For all Voronoi-cells, the integrated intra-cluster distance (squared) equals the error in relativistic energy associated with the local Voronoi-cell merge. The effects of merging particles into a cluster will always lead to energy loss (locally in that cell), but globally the error will become small because the loss of local momenta of the training vectors is counter-balanced by other clusters which carry part of the missing momentum and energy. \\

\noindent We have supplied a demonstration of this convergence property in energy/momentum conservation, of the k-means solution, in Section~\ref{sec:test_e_p}.

%----------------------------------------------
\subsubsection{Particle splitting}\label{sec:manifoldsplit}
An accurate method for splitting particles is also needed; when the particle number in a cell falls below some given threshold, an increase in phase space resolution becomes imperative --- even for physical reasons. 

\begin{figure}[!h]
\begin{center}
\includegraphics[width=.75\columnwidth]{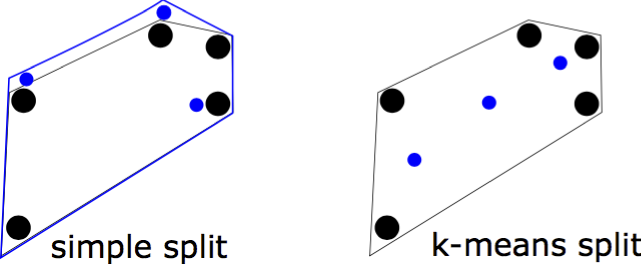}
\caption{Simple split \emph{vs} complex k-means based split, with re-distribution. Blue are the new particles, line designates boundary of a Voronoi cell. Particle size is arbitrary (not weighted in sketch). \emph{Left:} The simple split with a Gaussian (or other random) re-distribution is rather local and adds little phase space resolution. \emph{Right:} the more costly and complex k-means based splitting uses the phase space information already availble to place the newly splitted particles (codebook vectors) at maximal 6D distance from all other particles.}
\label{fig:split_simple_vs_complex}
\end{center}
\end{figure}

The problem of placing a large number of splitted particles in an optimal way, to make maximal use of the added CMP resolution in terms of information content, amounts to placing the splitted particles as far from their mother particles in phase space as permitted. This means much farther than a simple random splitting procedure as described above. We have sketched this situation in Figure~\ref{fig:split_simple_vs_complex}. 

A number of possibilities exist, many similar in approach, generally all based on a random or guided displacement in phase space of the mother and child particles. We add in this paper our novel approach of placing the newly split particles according a symmetry argument --- exploiting symmetries in the k-means based method --- to place the new particles according to the weighted k-means solution which works for splitting \emph{as well} as merging.

%--------
\subsubsection*{''On-top'' splitting; no re-distribution}
The intuitively simple ''on-top'' splitting into many new particles\cite{lapenta1,grasso1} is fast, exact, and guarantees perfect energy and momentum conservation at the time of splitting. Particles initially follow identical trajectories, after a single ''on-top'' split. Subsequently, as multiple splits are performed, particles attain unequal weights due to random selection of the initial codebook. Since a mother particle will carry a different weight than child and grandchild particles (which now instead carry equal weights), it will not in general - now - follow the same path anymore. As a result integration inaccuracies will develop, albeit extremely slowly. Such inaccuracies are generally ignorable over the course of an entire simulation. While a pure, exact, ''on-top'' split is always superior in performance and conservation of physics, such a pure ''on-top'' split, it does nothing to improve the fidelity of the simulation, only, it adds particles with no additional information content --- hardly a gain.

\subsubsection*{''On-top'' splitting; simple re-distribution}
A Gaussian (or even random) perturbation to particle pairs generated in the on-top split can safely be applied (e.g. \cite{may_silva_2014}), if it is desired to have particle pairs more quickly depart from exactly coinciding trajectories, for a coarse increase of phase space resolution.

We also tested this idea and can confirm that, even for perturbations (post-split) as large as ${\widetilde{\rr}}_{1,2}^{\prime} = {\widetilde{\rr}}_{1,2} + \widetilde{\bf{\delta}}\cdot{\widetilde{\rr}}_{1,2}~,$ ('prime' denotes perturbed particles, and $\widetilde{\bf{\delta}}$ a 6D Gaussian random variable and of order $10^{-2}$) energy and momentum were conserved virtually to machine precision, with $\Delta E \sim \mathcal{O}(10^{-6})$, see also Figure~\ref{fig:simple_split_tests}.

%--------
\subsubsection*{''On-manifold'' splitting; k-means re-distribution}
Interestingly, a complex k-means splitting procedure --- based on an argument of symmetry with particle merging --- proves to perform almost to the same accuracy in terms of conservation properties (also Fig.~\ref{fig:simple_split_tests}, right panel). Naturally, it is severely more expensive in terms of computational effort and memory consumption, yet, it re-distributes the additional CMP volume optimally spread over phase space, with maximally obtainable distance between the child (codebook vectors) and mother (training vectors) particles. We now describe this approach to splitting, symmetric with the k-means merging scheme.

We exploit now a symmetry of the k-means procedure to obtain an optimal re-distribution of split ''child'' particles with maximal spread on the phase space manifold. Increasing statistical resolution by adding particles can be achieved --- to the same numerical accuracy as merging --- by \emph{keeping} all training vectors, and \emph{adding} the codebook vectors

\begin{equation}
\{\widetilde{\rr}_{1},\ldots,\widetilde{\rr}_{M}\}_{\{s,tr\}} \rightarrow ~ \\
\{\widetilde{\rr}_{1},\ldots,\widetilde{\rr}_{M}\}_{\{s,tr\}}  +  \{\widetilde{\rr}_{1},\ldots,\widetilde{\rr}_{K}\}_{\{s,cb\}}~.
\end{equation}

%\begin{equation}
%\begin{aligned}
%\{\widetilde{\rr}_{1},\ldots,\widetilde{\rr}_{M}\}_{\{s,tr\}} ~~ & \rightarrow  & ~ \\
%\{\widetilde{\rr}_{1},\ldots,\widetilde{\rr}_{M}\}_{\{s,tr\}} ~~ & ~ + & \{\widetilde{\rr}_{1},\ldots,\widetilde{\rr}_{K}\}_{\{s,cb\}}~.
%\end{aligned}
%\end{equation}

Effectively, all new particles (codebook vectors), are placed \emph{precisely} on the 6D phase space manifold, but at positions \emph{different} from those of the original particles (training vectors), in 6D phase space. This amounts to a k-means phase space \emph{inflation} of type $M \rightarrow K \Rightarrow M+K$, in terms of number of CMPs. 

\begin{figure}[!h]
\begin{center}
\includegraphics[width=.8\columnwidth]{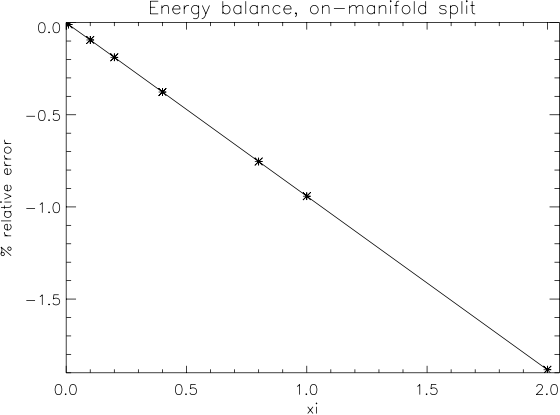}
\caption{Relative error in total energy for a single on-manifold split as function of the parameter $\xi$. For a split with $\xi=0.25$ which corresponds to equal sharing of weight between training vectors (original set) and the codebook (splitted particles), for $N_{cb} \equiv \sfrac{1}{3} N_{tr}$ the relative error is less than 0.3\%. For $\xi=0.001$ --- or 0.1\% transfer of the total weight --- the solution is still not as good as the Gaussian on-top split (not plotted here) in terms of energetics. The splitted particles are not contributing much to the dynamics, but they are still placed optimally under the objective of placing particles as far apart in phase space as possible. So, there is a trade-off between precision and weight transfer.}
\label{fig:split_energy_error}
\end{center}
\end{figure}

The only difference from merging is that we need to re-distribute the total weight of the original particle data, on \emph{both} training vectors and codebook vectors 
\begin{equation}
\sum_j^Kw_{cb} + \sum_i^Mw_{tr} = \sum_i^Mw^{\prime}_{tr}~,
\end{equation}
where now the $^\prime$ (prime) denotes values before performing k-means. Practically, the re-distribution of weights in our k-means based splitting scheme is done by sharing the weights between training vectors and their associated codebook vector in proportion to the training vectors' weights. The total amount of weight, $w_{cb}$, given to a codebook (cluster center), is the average value of those intra-cluster training vectors, 
\begin{equation}\label{eq:cbweights}
w_{cb} \equiv \xi \left\langle w_{cl}\right\rangle\frac{w_l}{w_{cb}}~~,~~\left\langle w_{cl}\right\rangle \equiv \frac{w_{cl}}{N_{cl}}
\end{equation}
is the mean weight in that cluster, and 
\begin{equation}
w_{cl} \equiv ~\sum_l^{N_{cl}}w_l,
\end{equation}
and $w_l$ are the individual weights of the $N_{cl}$ training vectors in that cluster. $\xi$ is a free parameter to either make the cluster receive less ($\xi<1$) or more ($\xi>1$) weight from the cluster members. The error introduced by the on-manifold split is linearly proportional to $\xi$, as seen from Figure~\ref{fig:split_energy_error}.

Splitting is, like merging, done under constraints of the mock edge preserving scheme, described in Section~\ref{sssec:convex}.

Particle splitting, which could have been simply $\mathcal{O}(K)$, now becomes a rather expensive --- as expensive as k-means for merging --- $\mathcal{O}(M \times K \times D \times i)$ once again. But; when taking into account an accelerated k-means algorithm (\citet{their_paper}), the computational feasibility of both splitting and merging is achieved; we can afford the extra care taken in undersampling (oversampling) particles phase space, for minimally (maximally) decreased (increased) statistical resolution.

%--------
\subsubsection*{''On-top'' \textit{vs} ''On-manifold'' splitting --- a check}\label{sec:split_compare}
We checked the performance of the ''on-manifold'' k-means based splitting/re-distribution scheme, and the simple ''on-top'' splitting/re-distribution scheme (with Gaussian perturbation of order 1\%), for a comparison of their ability to conserve energy over a series of splits. This test is comprised in the splitting stress test which is described further in Section~\ref{sec:stress_tests}.

In Figure~\ref{fig:simple_split_tests} (left panel) we plot the ratio of globally integrated particle energy, i.e.

\begin{equation}
\frac{E_{km}}{E_{ref}} = \frac{\sum_{i=1, s}^{N_{tr+cb}}e_{i,s}}{{\sum}_{i=1, s}^{N_{tr}} e_{i,s}} ~ ,
\end{equation}

with respect to a reference run where no splits are performed, as a function of time. All splitting methods conserve total energy to about $\mathcal{O}(10^{-3})$, for the duration of the test run (figure~\ref{fig:simple_split_tests}, right panel). \\

\noindent In summary, this counter-intuitive ''On-manifold'' approach to splitting does perform quite well. In terms of reproducing and conserving the physics (only), it is competitive, even if it falls short w.r.t. computational effort and memory consumption. We demonstrate the effectiveness of the K-Means based splitting method in the following sections on tests. Even if it is comparably expensive it could be desirable to employ the more expensive scheme in certain situations where maximal information content is to be extracted from the split particles, in a generic PIC code context. However, it is beyond the scope of this article to investigate the specifics of all physical scenarios where a complex split might be worthwhile choosing over a simple ''on-top'' split. Here we have merely argued that the symmetric operation, using k-means for splitting, performs well  while optimizing phase space spanning of the added statistical resolution (more particles). We also found for all other cases (not plotted here) that the 'on-manifold' splitting deviates from reference slower than the 'on-top' +Gaussian (1\% distortion in $\widetilde{\mathbf{r}}$) after many splits -- which hints that careful placement of added particles suppresses noise, which then grows with a Lyapunov exponent smaller than that of Gaussian noisy 'on-top' split. 

\noindent All tests throughout the remainder of this article (Section~\ref{sec:split_compare} primarily) have been performed with the k-means splitting method. Detailed work on ''on-top'' splitting is wide spread across the literature~\cite{lapenta1,lapenta2,lapenta3,scottmartin1,grasso1,may_silva_2014}. An exhaustively detailed comparison study of ''on-top'' \emph{vs} ''on-manifold'' splitting is beyond the scope of the present article.

%----------------------------------------------
\subsubsection{Contractive artifacts of K-Means}\label{sssec:convex}
\begin{figure}[!h]
\begin{center}
\includegraphics[width=.5\columnwidth]{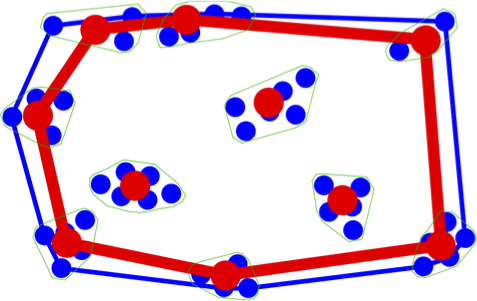}
\caption{Phase space volume contraction associated with k-means clustering in 2D. The convex portion of phase space (red perimeter line) spanned by the codebook clusters (red dots) will always be smaller than portion of phase space (blue perimeter line) spanned by the training vectors (blue dots). This effect is undesirable effect as it leads to edge depletion effects for $\rho_c(\rr)$ and $\JJ(\rr,\pp)$.}
\label{fig:convexity_sketch}
\end{center}
\end{figure}

The convex hull of a k-means solution will always \emph{contract} with respect to the original data set volume, i.e. 
\begin{equation}
\int_{\mathbf{\Omega}} d\mathbf{\Omega} f(\widetilde{\rr},t) < \int_{\mathbf{\Omega}^\prime} d\mathbf{\Omega}^\prime f^{\prime}(\widetilde{\rr},t)~,
\end{equation}   
except for the trivial case ($f^{\prime}==f$). Here $\mathbf{\Omega}, \mathbf{\Omega}^\prime \in \mathbb{R}^D$ are the convex hull bounding surfaces of the CMP density distribution in D dimensions before and after k-means compression. We have sketched this --- for PIC code applications undesirable property of k-means --- in Figure~\ref{fig:convexity_sketch} for D=2.

For PIC codes which are parallelized over computational processes via domain decomposition in real space ($\rr = \{r_x,r_y,r_z\}$) this is problematic because a given volume will experience edge artifacts in charge density, $\rho_c(\rr)$, and current density, $\JJ(\rr)$, namely a reduction of particles' contribution to those physical quantities. Domain decomposition is often employed in PIC codes, making such an edge preserving step indispensable.

To alleviate this problem we devise a simple mock edge preserving correction scheme, which is illustrated in Figure~\ref{fig:boundary_selection_procedure}. The idea is simply to let the clustered codebook solution approach the original training vector set on the domain boundaries. The boundary thickness is presently defined as equal to two cells of width. Is this way we ensure that edges are left untouched. In terms of computation, this leads to an extra iteration which we estimate at effort $\emph{O}(M \times K)$, thus not severe (yet not ignorable) in the total budget. Furthermore, the final number of codebook vectors will be slightly larger than the target value for large volumes and approach the original number of training vectors when the volume in question approaches PIC code cell size.

\begin{figure}[!h]
\begin{center}
\includegraphics[width=\columnwidth]{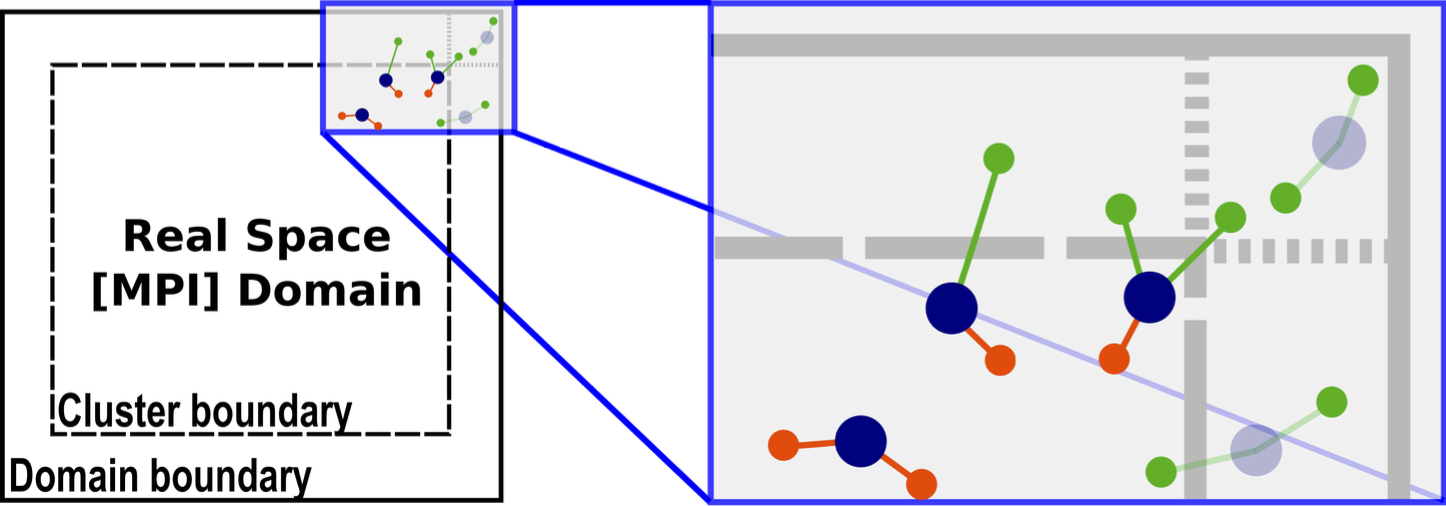}
\caption{Our mock edge preserving procedure, for spatially domain decomposed simulations. The procedure is applied only in those dimensions, where the convexity of the k-means leads to systematic edge effects. Shown in the inset on the right are: kept/omitted codebook vectors in dark blue/light blue, kept/deleted training vectors in green/red. See main text for further explanation.}
\label{fig:boundary_selection_procedure}
\end{center}
\end{figure}

This mock edge preserving procedure is simple; after having found a codebook solution on the entire domain (including the boundary region), the codebook vectors in the domain boundary are deleted, and the training vectors kept here instead. On the interior of the domain (excluding boundaries), the codebook is reduced if the cluster falls inside but has training vector members in the domain boundary. If a codebook vector resides on the interior and has \emph{all} training vector members on the interior as well, the codebook is kept as-is and the training vector members are deleted.

\begin{figure}[!h]
\begin{center}
\includegraphics[width=.75\columnwidth]{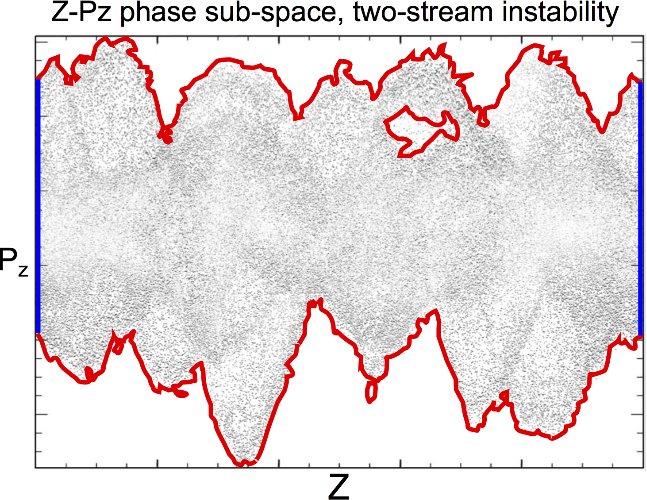}
\caption{2-dimensional $\{Z,P_z\}$ phase subspace. The contractive artifact of k-means in principle demands edge correction even in momentum space (red). One choice could be a contouring or swarm-based image segmentation procedure which could efficiently detect and maintain edges in momentum space. The real space boundaries (blue) are discussed in the section on our mock edge-correction for the (MPI) domain boundaries.l}
\label{fig:boundary_selection_procedure}
\end{center}
\end{figure}

Another issues concerning the contractive ''feature'' of k-means pertains to contraction in momentum space. While the domain boundaries are well defined in real space, $\mathbf{r}$, the clustering will contract momenta along approximate boundary contours in momentum space. This is sketched in figure~\ref{fig:pspace_contract} where we see that a streaming instability has a very complicated, sometimes even non-continuous, structure with holes, islands and wavy bounds. A mock edge preserving scheme which could alleviate momentum space contraction might be a contouring image segmentation procedure which could identify important contours in momentum space. This would significantly improve the overall ability to conserve not only charge distribution (real space egdes) but also energy and currents (momentum space ''edges'').

In Section~\ref{sec:barbara} we demonstrate that, despite its simplicity this edge preserving scheme manages to suppress edge-effects introduced by the contractive artifacts of k-means clustering, significantly. This can be appreciated from Figure~\ref{fig:step_edge_test}.

%=======================================================================
\section{Proof of concept: \texttt{'Barbara'} tests} \label{sec:barbara}
We proceed to demonstrate k-means based merging/splitting feasibility in terms of preservation of physics with heavily varying particle numbers. 

All tests in the remainder of this article have been performed using a slightly modified setup of a simple 2D3V relativistic two-stream simulation, used for tests of the \ppcode\cite{ppcode_paper}. A relativistic neutral electron-ion beam is streming through a neutral electron-ion background at $\Gamma_{beam}=3$, with density ratio $n_b/n_{bg}=1/3$. The dynamics are thought to be of relevance in cases such as Gamma-Ray-Burst afterglow shocks in a circumburst medium\cite{frederiksen2}.

Our reference \texttt{'barbara'} case in the present paper has grid size $N_{x,z}=128$, $N_y=1$, physical size $L_{x,z}=12\delta_e=3\delta_i$, $L_y=1.2\delta_e=0.3\delta_i$, $m_i/m_e=16$, beam Lorentz factor $\Gamma_{b}=3$, beam-to-background density $n_p/n_b = 1/3$, $\omega_{pe,0}\approx12$, $\delta_e=0.0856$, so $\delta_e/\Delta_x\approx8.6$. Time step $\Delta t=0.00391$, $t_{end}=10.0\approx120\omega_{pe}^{-1}\approx30\omega_{pi}^{-1}$, $N_p=30$ in the background and $N_b=10$ in the beam plasma per cell/species; a total of 80 particles/cell.  

The detailed reference simulation setup is not important for our tests; the only objective is to see how well we preserve the physics w.r.t. a reference case. Throughout this Section, the 'reference run' denotes the instance of \texttt{'barbara'} which is devoid of performing merging and splitting. 

Although our simulations are setup in a quasi-2D3V reduced dimensionality, this does not influence our k-means tests; particles still have a single cell's degree of freedom, even in the $Y$-coordinate. When we take into account the normalization of data space (see Section~\ref{sec:weighted_kmeans}), we will a have a truly 3D3V phase space manifold to work with. 

The simulations were all done on a 4x4 MPI domain decomposed geometry using simply the MPI processes as our k-means spatial domains. Still, domain sizes are not limited in any way, except for a lower bound on volume of a few cells in each spatial dimension. This is because the edge preserving scheme will make the solution approach the original phase space density for very small volumes of order \emph{a few} cells, $V_{kmeans} \equiv k\Delta_xl\Delta_ym\Delta_z$, where $\{k,l,m\} \rightarrow \{1,1,1\}$. 

We have verified the binary authenticity of successive reference runs, and that runs of \texttt{'barbara'}, using actual merging/splitting, were also binarily identical to the reference run -- up to the point of first k-means, of course.

%-------------------------------------------------------------------------
\subsection{Basic tests: energy-momentum conservation}\label{sec:test_e_p}
To demonstrate the energy-momentum conservation property of the k-means merging scheme, as discussed in Section~\ref{sec:e_p_cons}, we conducted a simple shell-based energy integration test. It shows that the error in energy is \emph{local} to the cluster at the Voronoi-cell center; it does not influence the global conservation properties of the k-means solution. 

The test was done for four instances of 'barbara' with varying particle number densities, at an early stage of evolution, when the background plasma can still be considered thermal and uniform. We tested a single species (electrons) of the background plasma.

Our test measures the difference in total energy between a pre- and post-merged solution. We measure the error in energy as we integrate the particles' energy inside successively expanding shells centered on the Voronoi-cell center, in real 3D space. These shells are indicated in Figure~\ref{fig:VoronoiDelaunay} (black dashed circles), where the plot now represents the real space clustering, rather than momentum space or general phase space. According to our argument (Section~\ref{sec:e_p_cons}) when considering a volume containing the surrounding Delaunay-cells on expanding shells, the error should become small.

\begin{figure}[!h]
\begin{center}
\includegraphics[width=\columnwidth]{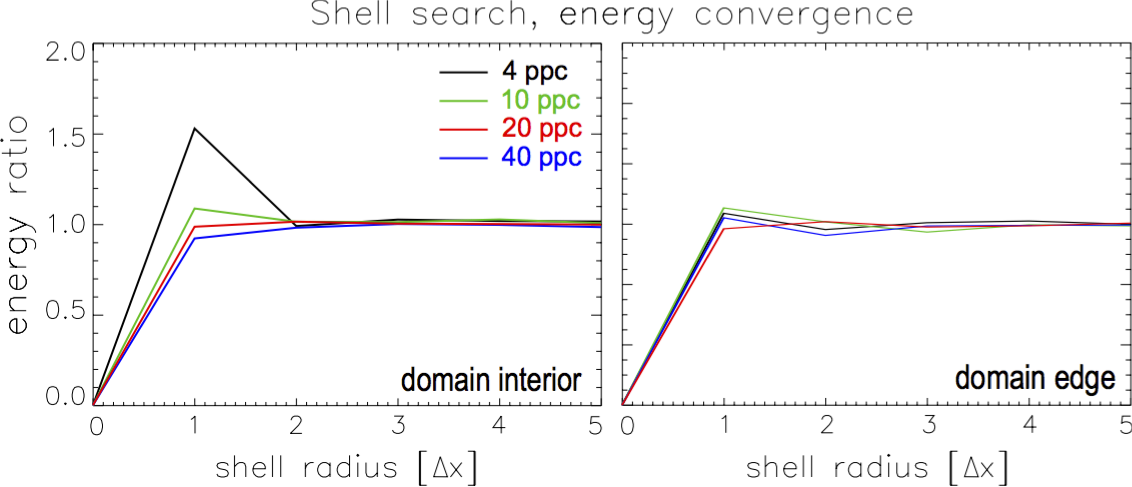}
\caption{Integration over shells of varying radius, around a Voronoi-cell center (cluster center) for energy conservation test, for four cases of initial particle densities. \emph{Upper}: the interior of the domain (where no edge correction is applied. \emph{Lower}: in the domain boundary, where edge correction is maximally applied. The edge correction provides slightly better conservation properties, of course, since the original solution is kept on the boundary. For $r_{sh} \sim 5\Delta x$ the edge case is about 1\% for all cases except case 4ppc ($\sim$5\%).}
\label{fig:shellsearch}
\end{center}
\end{figure}

Indeed, the energy conservation converges as expected when the neigboring Voronoi-cells (clusters) are included in the integrated energy. This is shown in Figure~\ref{fig:shellsearch}, where the energy ratio, $E_{tot,i}/E_{tot,f}$ is plotted for four cases of initial particle density, or 'particles-per-cell' ('ppc'). The energy is conserved as soon as the surrounding the dual set Delaunay-cells have been included in the energy calculation. This underlines the non-local nature of the solution, but also shows that the non-local error is actually rather almost-local, as it becomes small (and constant) at a radius of only 2 cell radii; the error is optimally small. The error will always bounded from below by the number of particles, the compression ratio and the convergence criterion set for k-means (the minimizing of distance measure, eqn.~\ref{eq:kmeansobjective}). 

The energy is conserved to about 5\% (for 4 ppc) when the full nearest neighbor Delaunay-cells are enclosed in the integration shell, $r_{sh} \sim 2\Delta x$. Energy is conserved slightly better for 10, 20, and 40 ppc at $r_{sh} \sim 2\Delta x$, and at $r_{sh} \sim 4\Delta x$, with values ranging from is 3\%-1\% for those latter three cases. The same conclusion is also reached for the momentum, conponent-wise. The same conclusion is reached for streaming particle species (which is just a similar k-means procedure in a mean Lorentz boosted frame). 

Energy and momentum conservation \emph{does}, however, become challenged on the k-means' domain boundaries (MPI domain bounds). This is our motivation for devising a mock edge preserving scheme (Section~\ref{sssec:convex} below). 

Our mock scheme is a quick solution, and can be further improved by using proper Delaunay-cell based selection of particles. Possibly including a window function smoother than a step function filter towards the edge, e.g. with a discontinuous step at $\Delta \mathbf{\widetilde{r}}$ could improve this edge correction even further. It is beyond the scope of this work to devise the perfect edge correction, but within scope to demonstrate the need for such a feature -- and its ability to mitigate edge effects.

%--------------------------------------------------------------------------------
\subsection{Basic tests: mock edge preserving scheme}\label{sec:edgepreservation}
As previously explained (section~\ref{sssec:convex}), the k-means procedure possesses an intrinsic and undesired property when it comes to preserving particle phase space structure in PIC simulations. Since any calculation exploiting arithmetic means will produce volumes smaller than the original one, a domain decomposed PIC simulation will suffer boundary effects in the domain decomposition dimensions. For our case of the \ppcode, this is in phase subspace $\widetilde{\rr}_{3D}=\{x,y,z\}$ (see also Section~\ref{sssec:convex}). In fact, it will even suffer this constraint in the domain \emph{non}-decomposed dimensions (here momentum phase subspace, $\widetilde{\rr}_{3V}=\{p_x,p_y,p_z\}$). 

The domain boundaries, in position (''physical'') space, are completely predictable. It makes sense to counter-balance convexity issues of the k-means based procedure based on a spatial filtering of the real space coordinate boundaries. We regard $\widetilde{\rr}_{3D}=\{x,y,z\}$ as correctable. Momentum subspace boundaries, $\widetilde{\rr}_{3V}=\{p_x,p_y,p_z\}$, are regarded as non-correctable since in any trivial -- they will generally not be regular. In the $p_z$-direction, however, things are not so predictable. We stress that we have \emph{not} decoupled 6D phase space reconstruction by this procedure, only, we have ensured the convergence of the compressed (or inflated) k-means solution in a subset of dimensions. We have introduced no decoherence between position and momentum at all by this edge-preserving correction. 

Two tests were performed to check the mock edge-preserving scheme performance 

\begin{description}
\item [Thermal cases] with little or no bulk flow which leads to little or no replenishment of domain boundaries, keeping the domain boundaries quasi-static in terms of phase space evolution.
\item [Streaming cases] the well evolved two-stream simulations would produce extreme replenishment of domain boundaries which will test the scheme's ability to render advection across boundaries transparent.
\end{description}

These two extreme dynamics are often realized in PIC codes, even simultaneously. 
\begin{figure}[!h]
\begin{center}
\includegraphics[width=\linewidth]{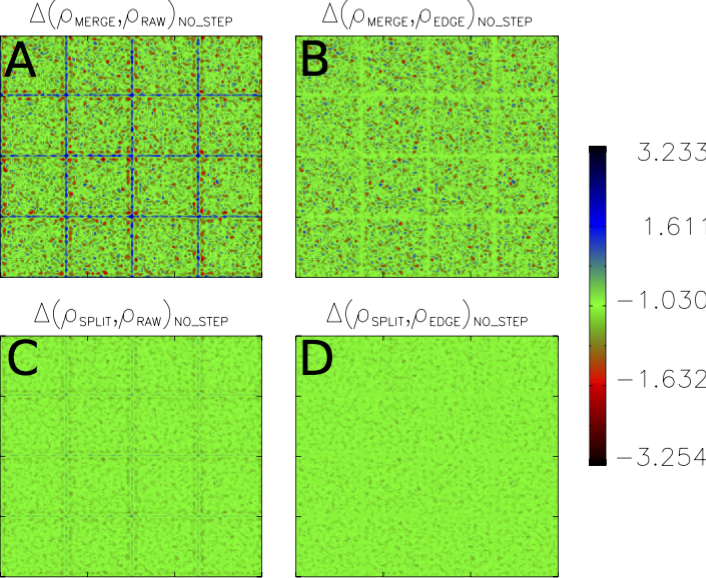}
\caption{Test of the mock-up edge preserving procedure. (A) (B): a single merge is performed with the usual PIC integration cycle completely omitted; the pure effect of the merge scheme is captured. Panel A(B) shows the merge without(with) edge-preservation effort, respectively. (C) and (D): same, but now a single SPLIT is performed instead.}
\label{fig:step_edge_test}
\end{center}
\end{figure}

The test results from the thermal case are shown in Figure~\ref{fig:step_edge_test}. We performed a single merge (split) --- and nothing else --- using the 'raw' non-corrected k-means algorithm, and a single merge (split) with our mock edge-preserving correction added to the k-means solver. The MPI domain boundaries are clearly visible. As expected the effect is more severe in the merging case since a split retains clusters with in-boundary particles whereas merging removes them. Still, even for merging, our edge-preserving correction yields significantly improved conservation properties. Not surprisingly both cases of splitting method were very accurate, to better than 0.1\% (also figure~\ref{fig:simple_split_tests}). Particle merging naturally showed a minor decay in the solution across the merge step since it is a case of lossy compression.  Still, the edge-preserving scheme forces the correct solution when approaching the real phase subspace boundaries, i.e. $f^{\prime}(x,y,z) \rightarrow f(x,y,z)$ for $\{x,y,z\} \rightarrow \{x,y,z\}_{\textrm{min,max}}$. The solution following several integrations would not decay rapidly. This result was reinforced by the streaming test case showing no appreciable discontinuities and ability to smooth away the boundary 'shadows' introduced by the contractive aspect of k-means arithmetic averaging.

%-------------------------------------------------------------------------
\subsection{Basic tests: pure merging \& splitting}\label{sec:stress_tests}

A stress test was performed to ascertain the quality and longevity of clustered solutions under what we defined as 'extreme' conditions. Multiple merges (or splits), only, were performed successively until the solution were no longer meaningful when comparing with the (constant particle number) reference run. After 400 iterations (33$\owpe$), we either only merged or only split the total particle number six times over the course of an additional 500 iterations (42$\owpe$).

The merging stress test successively attempted removal of 2/3 of the particles ($N_f=N_i/3$), for all species, on all MPI processes. Thus, from the first merge to the last merge, the number of particles would be $N_f/N_i = (1/3)^6 \approx 0.001$, had the fraction been exactly $\slfrac{1}{3}$. However, since we employed the mock edge preserving scheme (see Section~\ref{sssec:convex}) included which limits the boundaries to the true solution CMP number density, the remaining particle number fraction in each merging step was somewhat higher than 1/3. In fact the final-to-initial CMP total particle number was rather $N_f/N_i\approx 0.02$.

In a similar fashion, we conducted a k-means based splitting test (a less severe problem), which successively attempted to add 1/3 of the particles ($N_f=4N_i/3$), for all species, on all MPI processes, to the new solution. We emphasize our employment, here, of the expensive on-manifold k-means based splitting (see Section~\ref{sec:manifoldsplit}). Now, instead, the number of particles would have increased to $N_f/N_i = (4/3)^6 \approx 5.6$, had the added fraction been exactly 1/3. Again, the mock edge preserving scheme limited the true solution total number of CMPs to $N_f/N_i\approx4.7$. \\

\begin{figure}[!h]
\begin{center}
\includegraphics[width=\columnwidth]{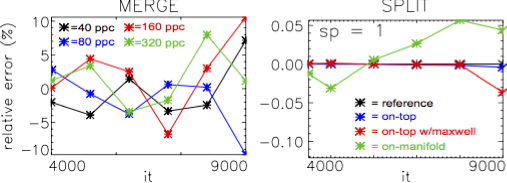}
\caption{Relative error for pure merging and splitting stress tests, with respect to a reference case. \emph{Left panel -- merging}, varying particle number. \emph{Right panel -- splitting}, varying method.}
\label{fig:simple_split_tests}
\end{center}
\end{figure}

Figure~\ref{fig:simple_split_tests} plots the energy conservation results for a selected species, for the merging/splitting stress tests (left/right panels).

During the first merge which reduces the total particle number to less than half, errors in total energy between 1\% and 3\% are introduced by the merge (left panel, fig.~\ref{fig:simple_split_tests}). A higher number of particles per cell ('ppc') leads to lower discrepancy although the variation is not a strong function of 'ppc'. Further, we find that the error grows slowly with successive merges, to a maximum of 10\%, only when the number of particles has been reduced by about 98\%.

While the ''on-top'' split is superior (right panel, fig.~\ref{fig:simple_split_tests}), it does not give way for improved fidelity. The ''on-manifold'' has larger error initially (as large as 0.1\% at times) but does provide a more full span of the phase space manifold. The test used $\xi=0.25$ (see also discussion of Equation~\ref{eq:cbweights}). The Maxwellian perturbed split population does no better than the ''on-manifold'', in some cases worse, for many successive splits and iterations. \\

\begin{figure}[!h]
\begin{center}
\includegraphics[width=\linewidth]{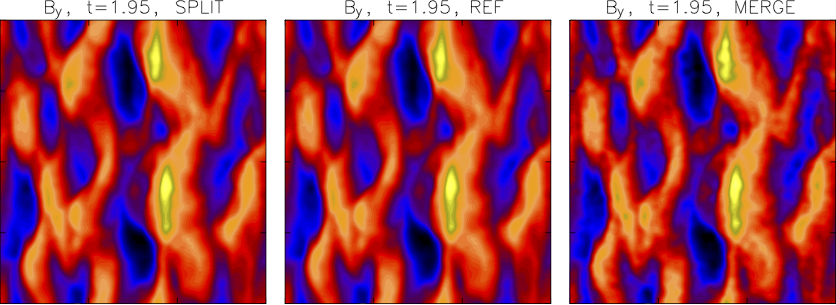}
\caption{Comparison of $B_y(x,z)$ for a hard stress test of merge (right panel) and split (left panel) with the reference case (middle panel), at time $t \sim 23 \owpe$. The number of particles has been merged (split) twice, into $N_f/N_i\approx0.24$ ($N_f/N_i\approx1.68$) over the course of 200 iterations.}
\label{fig:by_39_comp}
\end{center}
\end{figure}

\begin{figure}[!h]
\begin{center}
\includegraphics[width=\linewidth]{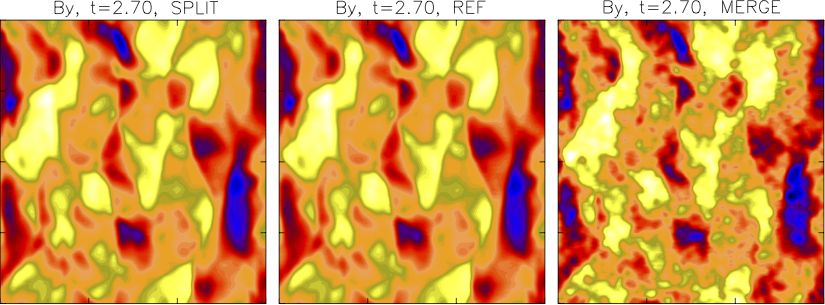}
\caption{Same as Figure~\ref{fig:by_39_comp}, but now, at time $t \sim 32 \owpe$ . The total number of particles has been merged (split) into $N_f/N_i\approx0.07$ ($N_f/N_i\approx2.8$) in four steps over the course of 400 iterations. The stress test has severely increased noise in the merge case.}
\label{fig:by_47_comp}
\end{center}
\end{figure}

The resulting evolutions of the merged and split cases are also worth comparing for a quantity which is solved for by particle-mesh interpolation and subsequent time integration, i.e. comparing Maxwell's source terms, $\JJ$ and $\rho_c$. In Figure~\ref{fig:by_39_comp} and Figure~\ref{fig:by_47_comp} we compare $\BB_y(x,z)$ for splitting (left panel), reference (middle panel) and merging (right panel) cases, for two splits (or merges) over 200 iterations, and four splits (or merges) over 400 iterations, respectively. All three panels are contoured on the same color scale, in each of the figures separately; they are directly comparable. Structure, phase, spectral properties are affected severely only after 4 merges, which introduces noise due to the decrease in particle numbers by $\sim$92\%.

Not surprisingly, the splitting produces an almost perfect match with respect to the reference case for all cases. Also as expected, the merged simulation shows increased levels of Poisson noise in the field as the number of CMPs is reduced. Still, after having removed more than 75\% of the original data set, and after several tens of iterations, the solution is still very good. Even after having removed more than 90\% of the particles over an additional 200 iterations, noise levels have risen considerably overall. The global evolution is well preserved. Although high-k noise has been introduced, global field structures are still well represented. We checked the Fourier spectra which show this behavior as well; high-k modes are rising during a merge stage, but the spectrum remains largely unaltered for low- and intermediate-k wavenumbers. 

After four merges, the total field energy has \emph{increased} by $\sim 5\%$, whereas the particle energy has \emph{decreased} by the same amount. This effect is also briefly mentioned in Section~\ref{sec:fullrun}, where the total energy budget is plotted for a full 'barbara' run, in Fig.~\ref{fig:diff_energy}. We interpret this counter-balance between fields and particles, as an increase in fields' noise on all scales due to the lower number of particles after a merge, from grid scale to systemwide scale. In future work it should be investigated whether this effect is robust; that would be a positive ''feature'' of the noise properties of the k-means global merging scheme.

This stress test qualitatively gauges the severity of compression parameters and resilience of compressed solutions. Much like image compression, we see that a reduction works well even for very high compression ratios, like for example compressing bitmapped images (\texttt{.bmp}) to Joint Photographic Experts Group images (\texttt{.jpg}), although the compression algorithms for images are likely more sophisticated in the latter case. 

A first, back-of-the-envelope quantification of validity of compression solutions hints that we should not compress (merge) successively more than 2 or 3 times without replenishing the CMP ensemble (either by advection or by production through collisions or simple statistical splitting). Furthermore, the compression ratio should not exceed about $\mathcal{R}_{merge}\equiv N_f/N_i \approx 2/3$ and not be performed successively within less than about a plasma-oscillation (10-20 iterations here). This estimate is based on empirical evaluation, and should be subjected to refined studies and analysis. \\

%-------------------------------------------------------------------------
\subsection{Basic tests: k-means, thermal distribution}\label{sec:stress_tests}
To test the k-means procedure's ability to retain a thermal ensemble, an alternating splitting-and-merging tests was performed. Using the \texttt{'barbara'} setup, with $\Gamma_b \equiv 1$ (no beam), we alternated between splitting and merging for 39 time steps, with $\Delta t = 0$, for a total of 19 splits and 20 merges, interwined.6 tests were run, with temperatures $T \in [10^{-1}, \ldots, 10^{-6}]$ (natural units, with $c\equiv1$) to check for sensitivity to momentum magnitude of the ensemble.

\begin{figure}[!h]
\begin{center}
\includegraphics[width=\linewidth]{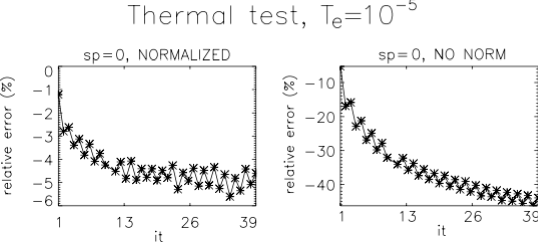}
\caption{Energy conservation thermal tests, plotted for species 0. Successive alternating merging (splitting) 20 (19) times, for a total of 39 k-means operations. Shown are cases comparing non-normalized and normalized data (see also section~\ref{sec:weighted_kmeans}. All other species show the same behavior.}
\label{fig:thermaltest}
\end{center}
\end{figure}

We plot the result concerning energy conservation in figure~\ref{fig:thermaltest}, for the case $T_e = T_i = 10^{-5}$ ($k_B\equiv 1$, $c\equiv 1$), calculated in nomalized data space (left panel). For reference, we plot also the energy ''conservation'' when calculating k-means in non-normalized data space (right panel), which underlines our claim that care should be taken not to ''disfavor'' (or underweigh) any dimension(s) in phase space, thus warranting our statement about normalizing the data. The plot also shows that merging conserves energy worse than splitting, as expected.

The thermal successive split/merge test employed the complex 'on-manifold' splitting scheme. Better energy conservation could be expected with a simple 'on-top' split since particles would simply cluster and re-fragment with no gain or loss in fidelity -- thus being less useful.

%-------------------------------------------
\subsection{Full scale automated merge/split test}\label{sec:fullrun}
We conducted two tests with the full scale automated MPI-domain based merging-splitting activated: a ''wide'' and ''narrow'' tolerance range test would decide how well, and how often splitting and merging should be employed. Again, the test is conducted on the 'barbara' case from above.

\begin{description}
\item [''Wide''] tolerance yielded splitting when, for any MPI domain, $N_p<N_{low} \equiv 0.667 N_{opt}$ and merging when $N_p>N_{high} \equiv 1.333 N_{opt}$
\item [''Narrow''] tolerance yielded splitting when, for any MPI domain, $N_p<N_{low} \equiv 0.9 N_{opt}$ and merging when $N_p>N_{high} \equiv 1.1 N_{opt}$. 
\end{description}

Due to the more restricted tolerance, the ''narrow'' test case yielded about twice as many splits and merges during the entire simulation, therefore also twice as many passes through the domain boundary edge-filtering. 

\begin{figure}[!h]
\begin{center}
\includegraphics[width=\linewidth]{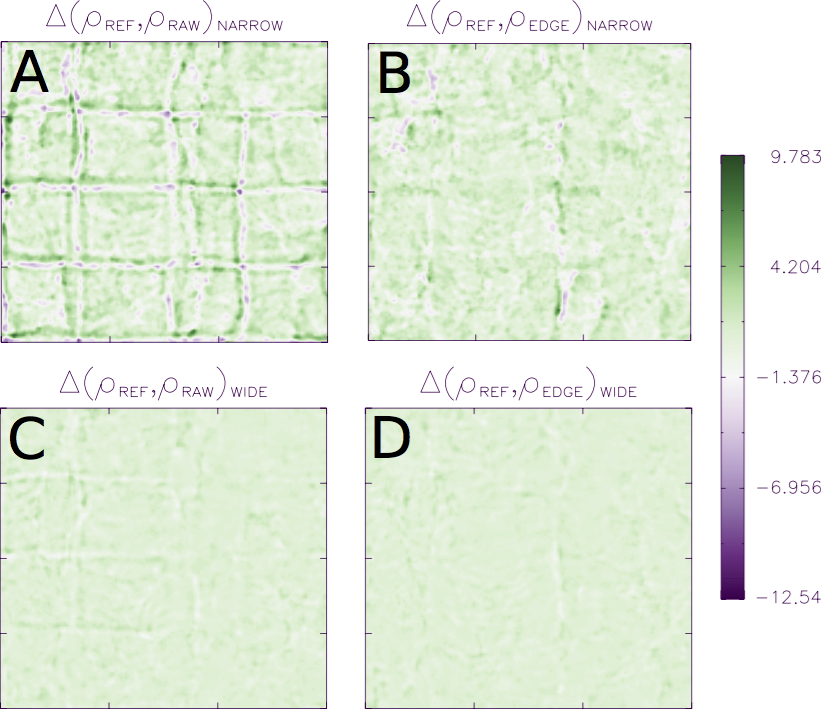}
\caption{Test of automatic merging and splitting. (A) and (B): procedure employing narrow tolerance range for the cases of ''raw'' and ''edge-preserving'' cases, respectively. (C) and (D): procedure employing wide tolerance range for the cases of ''raw'' and ''edge-preserving'' cases, respectively. The mock edge-preserving method is clearly superior.}
\label{fig:edge_preserve_test}
\end{center}
\end{figure}

We have plotted the differences w.r.t. reference when running raw, respectively edge-preserving, k-means in Figure~\ref{fig:edge_preserve_test}, for the ''narrow'' (panels 'A'=raw and 'B'=edge-preserving) respectively ''wide'' (panels 'C'=raw and 'D'=edge-preserving) tests. 

The effect of the more frequent splits/merges for the ''narrow'' case shows that traces of the MPI boundaries are visible in both the raw and edge-preserving cases. Yet, the quality of the edge-preserving scheme is still superior by a factor $\sim$2-5, i.e. 

\begin{equation}\label{eq:edgeerror}
\frac{\left[ \Delta \rho_{max} - \Delta \rho_{min} \right]_{RAW}}{\left[ \Delta \rho_{max} - \Delta \rho_{min} \right]_{EDGE}} \sim \rm{2-to-5}~.
\end{equation}

For the ''wide'' tolerance case, the more infrequent need for splits or merges yields improvement in the handling of edge effects at domain boundaries. Here, again, the edge-preserving scheme is justified, but now much better, namely by a factor 5-to-10 improvement (calculated as in equation~\ref{eq:edgeerror}. \\

\noindent For the remainder of this section we concentrate on the ''wide'' tolerance case. More than 2500 iterations, 100 merges, and 100 splits, were performed ($\sim$6 splits and $\sim$6 merges for each MPI domain). The splitting/merging kicks in at approximately the end of linear instability growth; this is expected since growth of current filaments, and charge separation results in concentrations of computational particles at that approximate time. 

\subsubsection*{Energy conservation}
\begin{figure}[!h]
\begin{center}
 \includegraphics[width=\columnwidth]{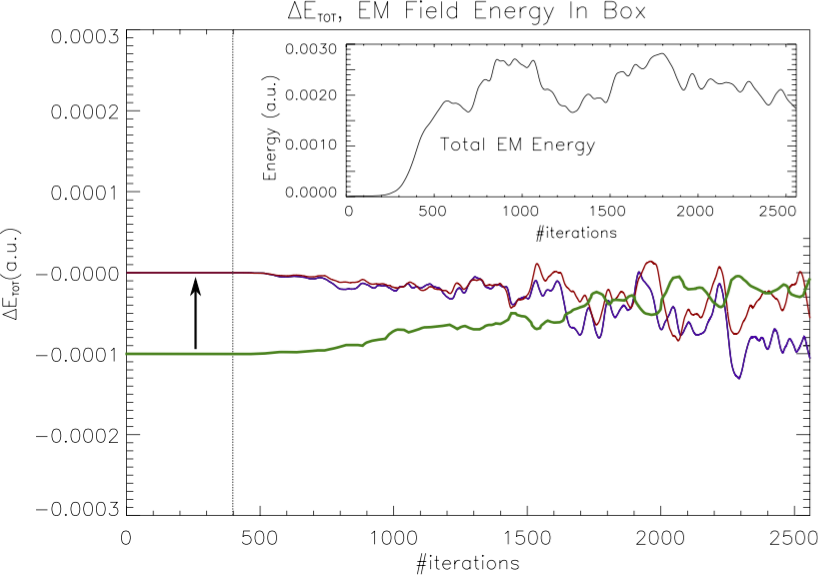}
\caption{Drifts in electromagnetic field energy (blue and red) and total negative particle energy (green, offset by -0.0001 for clarity) in the ''wide'' automated test case, using either a ''raw'' (blue) or an ''edge-preserving'' (red) methods (see section~\ref{sssec:convex} on edge preserving mock-up procedure). Inset: total EM energy in the simulation volume as function of iterations.}
\label{fig:diff_energy}
\end{center}
\end{figure}

Energy conservation and momentum conservation is striking, when we also look to the preservation of structures --- both spectral and spatial. In Figure~\ref{fig:diff_energy} is plotted the difference, $\Delta E(\textrm{ref,edge}) \equiv E_{\textrm{ref}}(t) - E_{\textrm{edge}}(t)$, $\Delta E(\textrm{ref,raw})$, in total electromagnetic field energy, and the \emph{negative} difference, $\Delta E(\textrm{edge,ref})$, in total particle energy, between the reference and ''wide'' test cases. 
Although the drift in EM energy is relatively large over time (about $1\cdot10^{-4}/2\cdot10^{-3}\sim0.05=5.0\%$), this energy drift is compensated by an anti-correlated drift in the total particle energy (see Figure~\ref{fig:diff_energy}). The drift in total particle energy is partially due to the convex artifacts of k-means, operating in momentum space -- which leads to artificial cooling, see also discussion in section on tests. The total energy deviates less than 0.5\% from reference in the case of wide tolerance range at the end of the simulation, after more than 200 merges and splits over more than 2000 iterations.

\subsubsection*{Momentum conservation}
\begin{figure}[!h]
\begin{center}
 \includegraphics[width=\columnwidth]{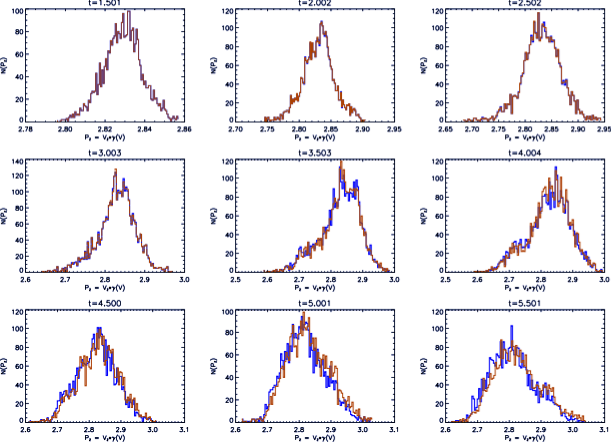}
\caption{Sample histogram of momentum space, ion beam species, along streaming direction, $\{P_z,N_p(P_z)\}$, for various times from just prior (upper left panel) to first merge, until about 1100 iterations ($\sim$48$\omega_{pe}^{-1}$) later.}
\label{fig:pz_hist}
\end{center}
\end{figure}

Likewise, we can study the total momentum evolution, but it would yield little new information. Rather, it makes sense to look at a ''critical'' component of the momentum; the ion beam momentum in the streaming direction. We have plotted the time evolution for this quantity in Figure~\ref{fig:pz_hist}. It is unnecessary to plot any other histograms in phase subspace since all directions are equally valuable in the k-means optimization procedure; they will show comparative accuracy.
There is a slight shift of the k-means treated runs (red curve) in the histograms as time progresses. We interpret this in connection with the conclusions concerning energy as a loss of energy transfer between particles and fields, which leads to a slower slow-down of the ion beam.

\subsubsection*{Particle weight distribution evolution}
\begin{figure}[!h]
\begin{center}
\includegraphics[width=\linewidth]{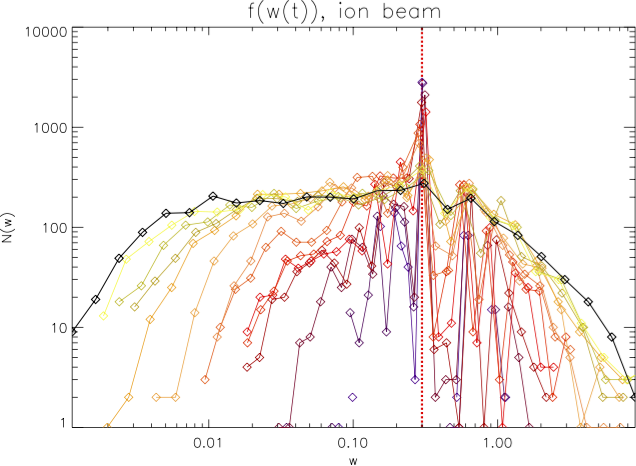}
\caption{Time evolution of particle weights, ion beam. Red dashes show the initial weight of all particles. Black curve final weights distribution at t=10.0 (\#it = 2553). Dark-to-light colored curves correspond to early-to-late time distributions, over time intervals of 128 iterations.}
\label{fig:weights_edge}
\end{center}
\end{figure}

Figure~\ref{fig:weights_edge} plots the evolution of particle weights (all particles) as the simulation progresses. From an initial constant weight, $w_{init}=0.3$, weights become distributed in a uniform manner over a wide range of values, as new generations of particles appear due to merging and splitting. 
This is desirable in terms of statistical evolution; phase space information is now also spread over a wide range in weights, and not only in phase space. We can more safely destroy particles at random without risking serious biasing effects on the physics in the process. For example for Monte Carlo modeling of collisional processes, a larger sampling space is available if more particles with smaller weights (and difeerent positions in 6D phase space) are available for the scattering process. Another advantage is that nearly empty, or at least very low-density, regions will also be more densely populated with particle phase space information. \\

\begin{figure}[!h]
\begin{center}
 \includegraphics[width=.75\columnwidth]{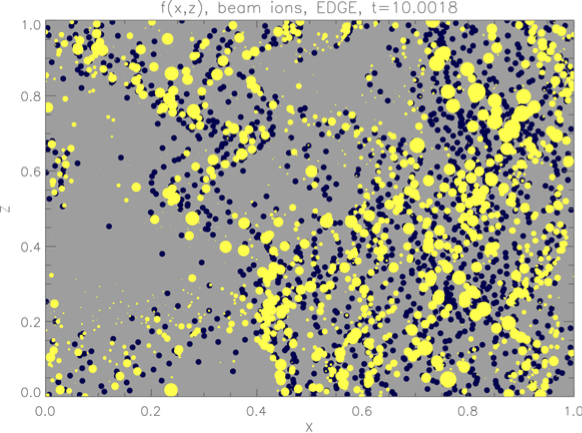}
\caption{Phase subspace $\{x,z,Log_{10}(w)\}$, for illustrative purposes. Yellow dots are late generation particles, size shows weight, while blue particles show concurrent reference run. Minor discrepancies in the local number density is due to a 1/100 stridden sampling --- and, to some degree, the collapse of phase space from 6D to 2D. This plot can be compared directly with Figures~\ref{fig:split_compare_field} and ~\ref{fig:weights_edge}}
\label{fig:particle_plot}
\end{center}
\end{figure}

\begin{figure}[!h]
\begin{center}
\includegraphics[width=\columnwidth]{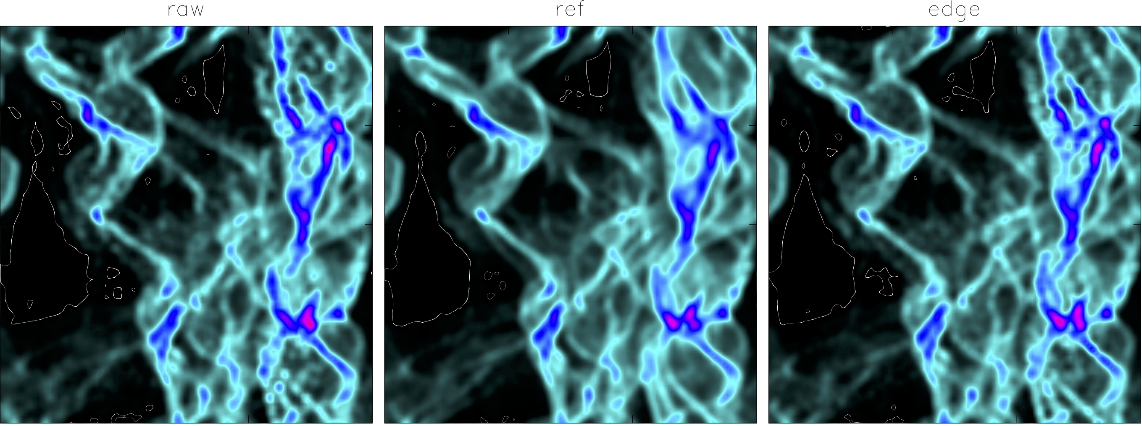}
\caption{Comparison between three runs of \texttt{''Barbara''} in the ''wide'' test case. Contour plots show beam ion density, during automatic merging and splitting on the wide tolerance interval. Approximately 100 merging and 100 splitting events occurred in total on the 16 MPI domains, for an average of about 12 splitting and merging events per domain. We have outlined the lowest level contour (thin white lines) to augment the finest level differences clearly. Number of contour levels is 256, $\text{max(}\rho_{b,i}\text{)}\approx 21.0$, $\text{min(}\rho_{b,i}\text{)} \approx 0.0$. \textit{Left panel:} raw k-means, no edge filtering.  \textit{Middle panel:} reference simulation (no k-means), \textit{Right panel:} edge filtered k-means. The edge-preserving method does perform slightly better, but the differences are very small.}.
\label{fig:split_compare_field}
\end{center}
\end{figure}

\noindent Figure~\ref{fig:split_compare_field} compares the beam ion density at the very last time step, after more than 2500 iterations and more than 100 splits and 100 merges. The result demonstrates that the solution stays stable for rather long times. There is a tendency for the particles to clump due to the frequent merges; the edge-preserving performs slightly better in avoiding clumps, and better preserves large scale structure and flow --- by a marginal measure.

%=====================================================================
\section{Discussion \& Concluding remarks}\label{sec:discussion}
It is important to realize that our k-means compression and inflation of CMP data in PIC codes is a \emph{global} method, which 'feels' all modes present in the volume at hand. Thus, the method will preserve all modes increasingly well for decreasing wavenumbers considered. This contrasts methods outlined during the Introduction, which conserve the physics \emph{locally}, rather. We briefly discuss here some obvious limitations of the k-means scheme. We then discuss how these limitations may not be too severe, considering a range of applications for which a global domain method can be utilized.

\subsubsection*{Applicability limitations on domain sizes}
There is a trade-off between the size of domains on which the method is deployed (here sizable MPI domains) and the speed at which the procedure can be executed. 

\begin{description}
\item [$V_{\textrm{kmeans}}\rightarrow V_{\textrm{cell}}$] In this case, our model approaches the full solution pre-merge/-split. We cannot go to cell sized domains due to the convexity issue, which then becomes either detrimental to conservation of the physics, or becomes nullified due to retainment of the full training vector set in question. 
\item [$V_{\textrm{kmeans}}\rightarrow V_{\textrm{total}}$] In this case, the edge effects approaches a negligible contribution. We \emph{can} encompass the entire domain; even though super-linear scaling becomes prohibitive for performance this is only a practical limitation, not a physical one. Also, the compression ratio parameter is relieved of any restrictions imposed by convexity in this limit.
\end{description}

While the latter limit can be remedied by accelerated algorithms or hardware acceleration (see below), the former cannot. We are bound by a lower limit on domain volumes to obtain a reasonable compression factor. The domain volume boundary thickness, $W_{\rm{bound}}$, must be small compared with the domain interior, along all dimensions. $W_{bound}\equiv1\dx<<L_x=32\dx$ (\emph{per} MPI domain) in our study above. Still, a domain volume granularity leading to $W_{bound} = 1\dx \lesssim L_x \equiv [a~few]\cdot\dx$ should be possible. 

\subsubsection*{Demands for memory}
A serious drawback of the current implementation is memory overhead; worst case, we need simultaneous storage for the codebook, as well as training-to-codebook vector mapping of particle IDs, and count of training vectors-\emph{per}-codebook vector. The total overhead then becomes $(7[real] + \slfrac{7}{3}[integer])N_{opt}$, which should be compared with the normal need for a PIC representation (when not employing merging/splitting) of simply $7[real] N_{opt}$ particles: an overhead which doubles the memory need. 

Careful implementation and re-use of allocated space can reduce the need for memory, but at present we see no way to remove the need for storage of the full codebook vector data set for the 'k-means' sceheme. This could improve in the future. \\

\noindent One quite obvious way to remove the problem of overhead in connection with the codebook construction might be to replace the weighted 'k-means' step with a weigthed 'k-medoids' scheme. Training vectors are then taken as codebook centers, thus re-cycling previously allocated memory in an elegant and efficient way. We are currently investigating whether 'k-medoids' will also perform adequately with respect to the physics --- this is not given \emph{a priori}. It is of similar computational complexity as 'k-means'.

\subsubsection*{Acceleration of K-Means clustering}
The standard Lloyd's k-means is too slow, even prohibitively so. We cannot obtain a process which is faster than about $\mathcal{O}(M \times K \times D \times i)$, with $D$ the dimensionality (which is 6 for 3D3V) and $i$ is the number of iterations to convergence. For our test case, the time spent in k-means calculations exceeded the entire simulation time by several factors. An accelerated method is clearly needed. To gain sufficient speed in the computation, we need an approximate factor of 40 in speed-up. This is only possible using more efficient algorithms or more efficient hardware, for the same problem size. \\

\noindent In a separate project, we have investigated acceleration by introducing a KD-Tree method, which seems promising when keeping calculations on CPU. Hardware acceleration is also an option, since Lloyd's k-means is ''embarrasingly parallel''; a GPGPU kernel (in OpenCL) was implemented, with considerable speed-up, on-core. However, the copying of data to-and-fro the GPGPU is costly, and only for large data sets does the hardware accelerated method become feasible. The question of algorithmically accelerated \emph{versus} hardware accelerated k-means will be treated in a subsequent publication (\citet{their_paper}, \emph{in prep}). 

For our \texttt{''BARBARA''} test, a typical accelerated k-means step executed about as fast as a typical simulation timestep, in the case of the KD-Tree algorithm, making the procedure competitive for several important applications (see Introduction). The GPGPU performed comparably, in some cases better than KD-Tree, but has only been benchmarked in \emph{stand-alone} tests (on large 6D particle data sets).\\

\noindent With this article we have given an account for a global k-means based phase space compression method. For PIC codes we have evaluated the \emph{physical} conservation properties without a glance to \emph{computational} effort. In a separate publication (\citet{their_paper}, \emph{in prep}) the question of computational feasibility is being addressed in two different ways; by \emph{algorithmic} acceleration (KD-Tree algorithm optimization with MPI+OpenMP support), and by \emph{hardware} acceleration (Lloyd's brute force on GPGPU). \\

\noindent For a small laser-plasma interaction case, with electron bunching (production run example; see Figure~\ref{fig:laserplasma_interaction}), we tested the KD-Tree accelerated k-means for 2, 4 and 8 OMP threads per MPI domain (total of 16-by-\{2, 4, 8\} threads).

\begin{equation}
\begin{array}{c|c|c|c|c}\label{tab:runtime}
 \texttt{OMP\_NUM\_THREADS} & 2 & 4 & 8 & \rm{Ref} \\\hline 
 T_{end}~[s]~\rm{(wall)} & 5296 & 5319 & 5782 & 6308
\end{array}
\end{equation}

Runtimes were compared with a reference run where, again, no particle control were active. For these cases, the runtimes are given in table~\ref{tab:runtimes}, where we see that the $\texttt{OMP\_NUM\_THREADS=8}$ performed most favorably, giving a reduction in runtime of about 15\%. This speed-up will increase with increasing simulation size, therefore increasing particle number density constrasts. The quality of the resulting structures/energy conservation has been verified, and matches very well the 'barbara' results quoted here.

Both methods, KD-Tree and GPGPU (latter not quoted here) acceleration, are indeed affirmative towards using our k-means procedure for realistic problems.

%------------------------------------------
\subsubsection*{Generality of K-Means Merging \& Splitting}
Particle phase space compression and inflation is not limited to electromagnetic PIC codes. Any system which can be modeled by a discretized particle distribution function, $f(\widetilde{\rr}(t),t)$, in a D-dimensional space ($\widetilde{\rr}(t)\in\mathbb{R}^D$) can be manipulated by the k-means optimization scheme described in this article. This means that
\begin{enumerate}
\item the number of particles must be high enough to consider the particle population(s) as approximating a continuous distribution within a given volume selected for merging, and
\item that the intrinsic noise in the non-reduced solution must outweigh the noise introduced by the reduction of the particle data set. 
\end{enumerate}
The specific values for these constraints are of course problem dependent, the determination of which are beyond the scope of this paper. This question is deferred to future work.

\begin{figure}[!h]
\begin{center}
\includegraphics[width=.75\columnwidth]{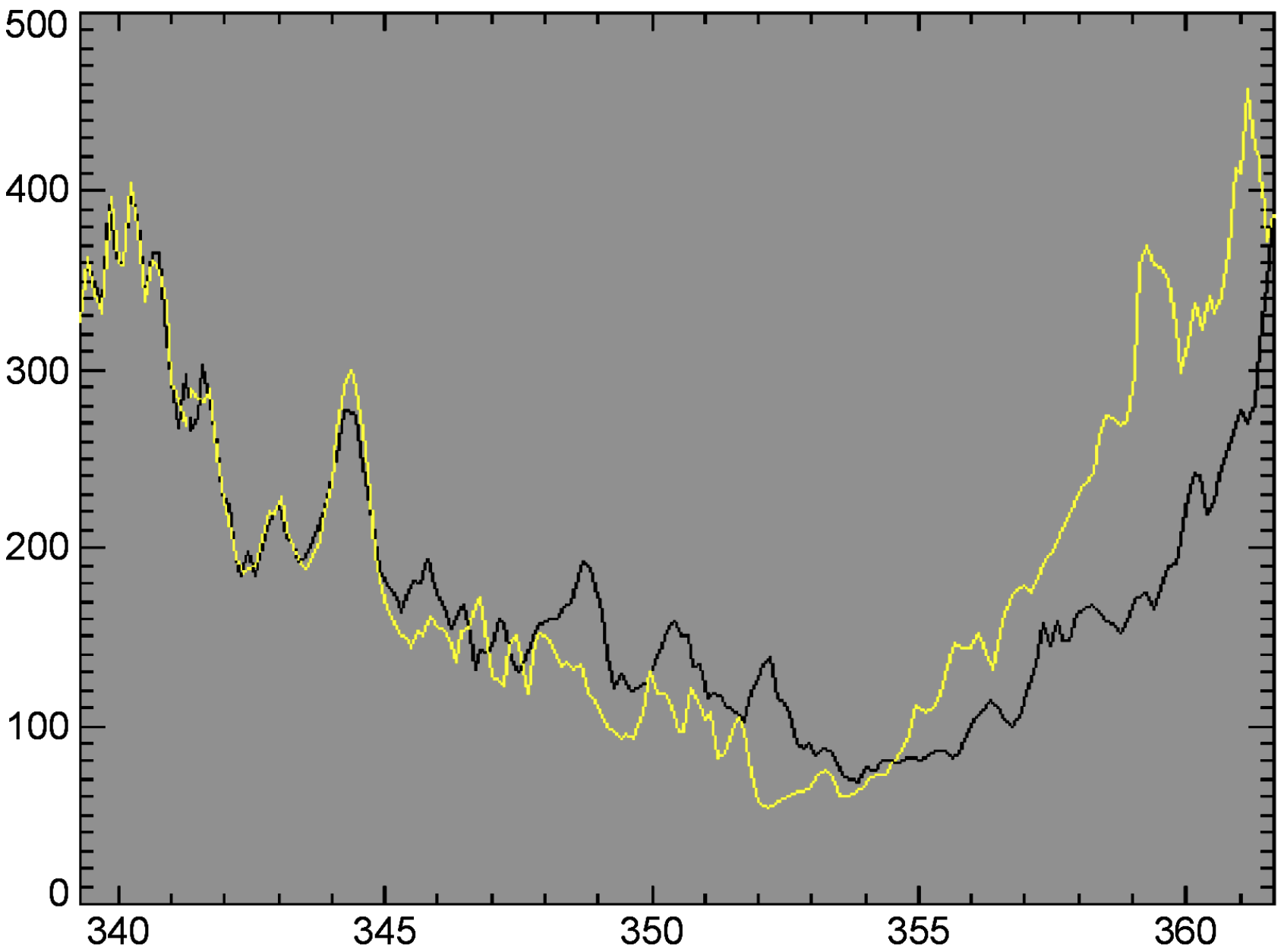}
\caption{Global maximum CMP number density, $\mathtt{max[}N_{\textrm{cell}}(t)\mathtt{]}$ in a stand-alone test of the k-means merging procedure on a PIC code used for simulating accretion of planetesimals in planet formation. Black: original particle set (2.4 million CMPs). Blue: reduced set (1.2 million CMPs).}
\label{fig:johansen}
\end{center}
\end{figure} 

Nonetheless, to give an example here, we have verified the method in one other case of a particle-in-cell based code, by Johansen \emph{et al.}\cite{johansen1,johansen2} which has shown promising results as well. At a well evolved point in time, in a simulation of the formation of streaming instabilities in a proto-planetary disk, a total of 2.4 million particles were merged into 1.2 million, and the simulation was restarted with the reduced codebook solution. Figure~\ref{fig:johansen} shows a comparison between a reference simulation (black) and the compressed phase space simulation (yellow). The quantity plotted is the (global) maximum CMP number density, $\mathtt{max[}N_{\textrm{cell}}(t)\mathtt{]}$, which is a very sensitive measure. Coherence is almost perfect after 25 iterations and still reasonable after almost 100, even for a reduction by half the particle ensemble. The instance of our k-means algorithm used in this stand-alone test was an early stage implementation, originally employing the Linde-Buzo-Gray algorithm\cite{lindebuzogray1}, rather than Lloyd-Forgy's. Also, we did not employ our edge-preserving procedure. The agreement plotted here is likely to improve in the future.

%-----------------------------
\subsection*{Concluding Remark}
Without the ability to make a direct comparison by benchmark, it is difficult to assess the range of validity, conservation abilities, computational cost, and possible domain boundary constraints, of the methods mentioned in the Introduction. When we consider the problem from a global PIC domain perspective, the issue of memory overhead associated with performing an analytical match of charge and current densities (without the tensor conservation) could possibly approach that associated with our k-means procedure when $M \lesssim K$. We therefore encourage a comparison of merging methods, in particular between more recent methods \citep{grasso1,scottmartin1}, and the statistical k-means based optimization scheme introduced in the present article. 

As kindly pointed to also by one of our reviewers, k-means sometimes tends to create clusters that are more uniform than the original distribution~\cite{xiong1}. This may challenge clustering-based merging algorithms when considering anisotropic or asymmetric distribution functions. Therefore the bump-on-tail problem\footnote{We thank one of our referees for pointing to this reference and the thorough review} could be included in a suite of common benchmarks for future comparisons of merging algorithms (see also Concluding remark, sec.~\ref{sec:discussion}). \\

\noindent The clustering merge/split code used for this publication may be requested by emailing the author\footnote{Proper citation of this article (or co-authorship) should follow application, transcript or re-engineering of the acquired code.}.

%----------------------------------------------------------------------
\section*{Acknowledgments}
%\begin{acknowledgments}
JTF is grateful to Mordechai Butrashvily and Luk\'{a}\v{s} Mal\'{y} for their work leading to the software and hardware accelerated versions of k-means (via the 'Summer of PRACE' exchange program, 2013). JTF is thankful for early discussions with Shahab Fatemi, on straightforward subgridding methods, and acknowledges \AA . Nordlund, K. Galsgaard and T. Haugb\o lle, for discussions concerning integration of the k-means framework into the ~\ppcode. Research leading to this article was initiated during the EU-FP7 Collaborative Project 'SWIFF' (\url{http://www.swiff.eu}). Simulations leading to the result presented in Figure~\ref{fig:laserplasma_interaction}, were performed~\cite{beck1} using the PRACE Research Infrastructure resource FERMI based in Italiy at CINECA. The research leading to these results has received funding from the European Research Council under the European Union?s Seventh Framework
Programme (FP/2007-2013) under ERC grant agreement 306614 (MEP). JTF and MEP acknowledge support from the Young Investigator Programme of the Villum Foundation.
%\end{acknowledgments}

%----------------------------------------------------------------------
%\bibstyle{pop}
 \bibliographystyle{chicagoa}
 \bibliography{paper}

%----------------------------------------------------------------------
\end{document}